\documentclass[reprint,superscriptaddress,amssymb, amsmath, aps, showpacs, footinbib,  prb]{revtex4-1}
\pdfoutput=1
\usepackage{graphicx}

\newcommand{\madCr}{V_M^{\text{Cr}}}
\newcommand{\madO}{V_M^{\text{O}}}

\begin{document}

\title{Importance of tetrahedral coordination for high-valent transition metal oxides: YCrO$_4$ as a model system}

\author{A. A. Tsirlin}
\affiliation{Max Planck Institute for Chemical Physics of Solids, N\"{o}thnitzer
Str. 40, 01187 Dresden, Germany}
\affiliation{National Institute of Chemical Physics and Biophysics, 12618 Tallinn, Estonia}

\author{M. G. Rabie}
\affiliation{Max Planck Institute for Chemical Physics of Solids, N\"{o}thnitzer
Str. 40, 01187 Dresden, Germany}
\affiliation{Departamento Quimica Inorganica, Facultad de Ciencas Quimicas,
Universidad Complutense de Madrid, 28040 Madrid, Spain}
\affiliation{II. Physikalisches Institut, Universit\"at zu K\"oln, Z\"ulplicher Str. 77, 50937 K\"oln, Germany}

\author{A. Efimenko}
\affiliation{Max Planck Institute for Chemical Physics of Solids, N\"{o}thnitzer
Str. 40, 01187 Dresden, Germany}
\affiliation{II. Physikalisches Institut, Universit\"at zu K\"oln, Z\"ulplicher Str. 77, 50937 K\"oln, Germany}

\author{Z. Hu}
\affiliation{Max Planck Institute for Chemical Physics of Solids, N\"{o}thnitzer
Str. 40, 01187 Dresden, Germany}

\author{R. Saez-Puche}
\affiliation{Departamento Quimica Inorganica, Facultad de Ciencas Quimicas,
Universidad Complutense de Madrid, 28040 Madrid, Spain}

\author{L. H. Tjeng}
\affiliation{Max Planck Institute for Chemical Physics of Solids, N\"{o}thnitzer
Str. 40, 01187 Dresden, Germany}

\begin{abstract}
We have investigated the electronic structure of the high oxidation state material YCrO$_4$ within the framework of the Zaanen-Sawatzky-Allen phase diagram. While Cr$^{4+}$-based compounds like SrCrO$_3$/CaCrO$_3$ and CrO$_2$ can be classified as small-gap or metallic negative-charge-transfer systems, we find using photoelectron spectroscopy that YCrO$_4$ is a robust insulator despite the fact that its Cr ions have an even higher formal valence state of 5+. We reveal using band structure calculations that the tetrahedral coordination of the Cr$^{5+}$ ions in YCrO$_4$ plays a decisive role, namely to diminish the bonding of the Cr $3d$ states with the top of the O $2p$ valence band. This finding not only explains why the charge-transfer energy remains effectively positive and the material stable, but also opens up a new route to create doped carriers with symmetries different from those of other transition-metal ions.
\end{abstract}

\maketitle

The nature of the lowest-energy charge excitations in strongly correlated transition-metal compounds is largely determined by the magnitude of the Coulomb energy $U$ at the transition-metal $d$ shell and that of the charge-transfer energy $\Delta$ needed to transfer an electron from the $d$ shell to the ligand $p$ states. As pointed out by Zaanen, Sawatzky, and Allen (ZSA),\cite{zaanen1985} systems with sufficiently large $U$ and $\Delta$ values are magnetic insulators, and can be classified as either Mott-Hubbard insulators (the lowest-energy inter-site charge excitation is between the $d$ ions) or charge-transfer insulators (the excitation is between the $d$ and $p$ ions). Compounds with small or close-to-negative $\Delta$ values can be metallic with $p$-type charge carriers, and CrO$_2$\cite{korotin1998,stagarescu2000,huang2003} as well as Cr$^{4+}$ compounds in general\cite{ortega2007,lee2009,komarek2008,streltsov2008,komarek2011} are representative for such a case.

One may expect that increasing the Cr valence further to 5+ could result in a truly negative-$\Delta$ system. This idea is based on a simple argument that $\Delta_{n+1}$ is given by $\Delta_n-U$, where $n$ enumerates the valence.\cite{zaanen1985} Surprisingly, YCr$^{5+}$O$_4$ and related RCrO$_4$ compounds with rare-earth R$^{3+}$ ions seem to be insulators according to their deep green color.\cite{jimenez2000,aoki2001,rcro4-2003,long2007,errandonea2011,qiao2013} This insulating behavior becomes even more puzzling when a band gap analysis, as developed by Torrance,\cite{torrance1991} is carried out to include the actual crystal structure effects on the Madelung potentials (Table~\ref{tab:madelung}). We find that the charge-transfer energy should be negative enough to stabilize a metallic state in YCrO$_4$. Our analysis should also be applicable to Sr$_3$Cr$_2$O$_8$ and Ba$_3$Cr$_2$O$_8$, which are insulators as well.\cite{radtke2010,singh2007,aczel2009,kofu2009,wang2011}

To resolve this issue, we have performed a photoelectron spectroscopy study on the valence band of YCrO$_4$ and combined it with \textit{ab-initio} and parameterized band-structure calculations. Experimental results evidence that YCrO$_4$ is a robust insulator. From the calculations, we discover that the tetrahedral coordination of the Cr$^{5+}$ ions plays a crucial role in diminishing the bonding of the Cr $3d$ states with the top of the O $2p$ valence band as to make the charge-transfer energy positive, to allow for the formation of a large band gap, and to keep the compound stable. Additionally, this lack of bonding with the top of the valence band opens new ways to create charge carriers by doping. Such carriers should have a high oxygen character with a symmetry decoupled from that of the transition-metal ions, and possibly generate unexpected phenomena.


\begin{figure}
\includegraphics{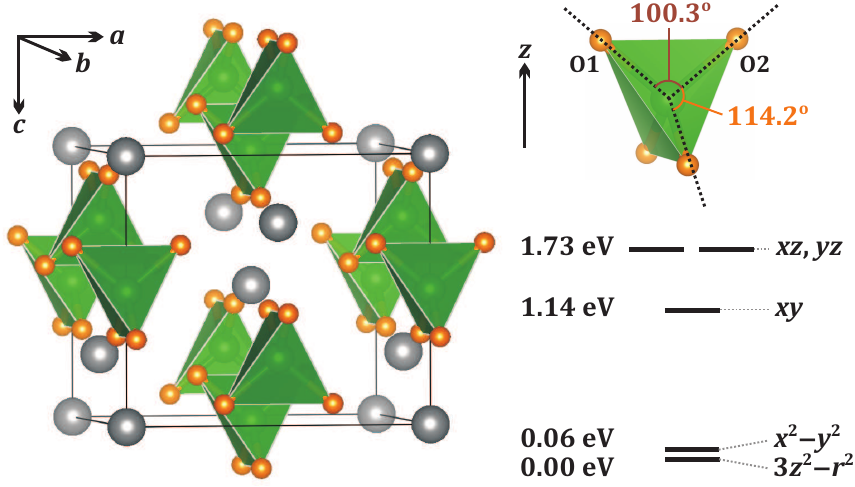}
\caption{\label{fig:str}
(Color online) Left panel: Crystal structure of YCrO$_4$ with isolated Cr$^{5+}$O$_4$ tetrahedra. Right panel: weak distortion of the tetrahedra and the crystal-field levels with two states of $e$ symmetry ($x^2-y^2$, $3z^2-r^2$) lying below three states of $t_2$ symmetry ($xy,xz,yz$). The energies are given according to LDA with respect to the lowest-lying $3z^2-r^2$ state.
}
\end{figure}
We focus on the tetragonal zircon-type polymorph of YCrO$_4$, for which well-characterized bulk samples are available. This compound develops a ferromagnetic order below $T_C=9$~K\cite{long2007} and its crystal structure (Fig.~\ref{fig:str}) comprises isolated Cr$^{5+}$O$_4$ tetrahedra and Y$^{3+}$O$_8$ bisdisphenoids zig-zag chains. We calculated the Madelung potentials at the O and Cr sites using the \texttt{TETR} utility.\cite{tetr} The potentials are listed in Table~\ref{tab:madelung} together with those of NaCrO$_3$, SrCrO$_3$, and LaCrO$_3$ as reference systems for Cr$^{5+}$, Cr$^{4+}$, and Cr$^{3+}$ materials, respectively. The calculations for the latter 113 perovskite-type compounds were all done in the idealized cubic crystal structure of LaCrO$_3$ with the same lattice parameter of $a=3.88$~\r A.

\begin{table}
\caption{\label{tab:madelung}
Madelung potentials $V_M$ of Cr and O, ionization potentials $I_v^{\text{Cr}}$ for Cr$^{v+}$, relevant Cr--O distances $d_{\text{Cr--O}}$, and the resulting effective charge-transfer parameters $\Delta_0$ according to Torrance.\cite{torrance1991}
}
\begin{ruledtabular}
\begin{tabular}{cccccr}
            & $\madCr$ & $\madO$ & $I_v^{\text{Cr}}$ & $d_{\text{Cr--O}}$ & $\Delta_0$ \\
            &   (V)    &   (V)   &     (eV)          &       (\r A)       &   (eV)     \\
  LaCrO$_3$ & $-38.4$  &  22.1   &    31.0           &       1.94         & 14.5       \\
  SrCrO$_3$ & $-46.0$  &  24.0   &    49.1           &       1.94         & 5.6        \\
  NaCrO$_3$ & $-53.5$  &  25.8   &    69.3           &       1.94         & $-5.3$     \\
  YCrO$_4$  & $-54.9$  &  26.8   &    69.3           &       1.66         & $-4.2$     \\
\end{tabular}
\end{ruledtabular}
\end{table}
Torrance~\cite{torrance1991} introduced an effective charge-transfer parameter $\Delta_0$ that combines different energy terms related to the electron transfer from the $d$ shell of Cr to the $p$ shell of O:
\begin{equation}
  \Delta_0=e(\madO-\madCr)+A^{\text{O}^-}-I_v^{\text{Cr}}-e^2/d_{\text{Cr--O}},
\end{equation}
where $V_M$ are Madelung potentials, $A^{\text{O}^-}=-7.7$~eV is the electron affinity for the O$^-$ ion (equal to the ionization potential of O$^{2-}$), $I_v$ is the ionization potential of Cr$^{v+}$, and $d_{\text{Cr--O}}$ is the Cr--O distance that determines the energy of an excitonic excitation in the ionic model. Torrance~\cite{torrance1991} found empirically that for a wide range of oxides the optical gap follows roughly the relation $E_{\text{gap}}\approx\Delta_0-10$~eV. The calculated $\Delta_0$ values would then suggest that, according to this empirical rule, LaCrO$_3$ should be an insulator, while SrCrO$_3$, NaCrO$_3$, and YCrO$_4$ should be metallic. This approach gives consistent results for LaCrO$_3$ (insulator), SrCrO$_3$ (metal~\cite{ortega2007,komarek2011}), and probably also for NaCrO$_3$ (not available so far), but apparently not for YCrO$_4$ as the latter material seems to be an insulator.\cite{long2007}

\begin{figure}
\includegraphics{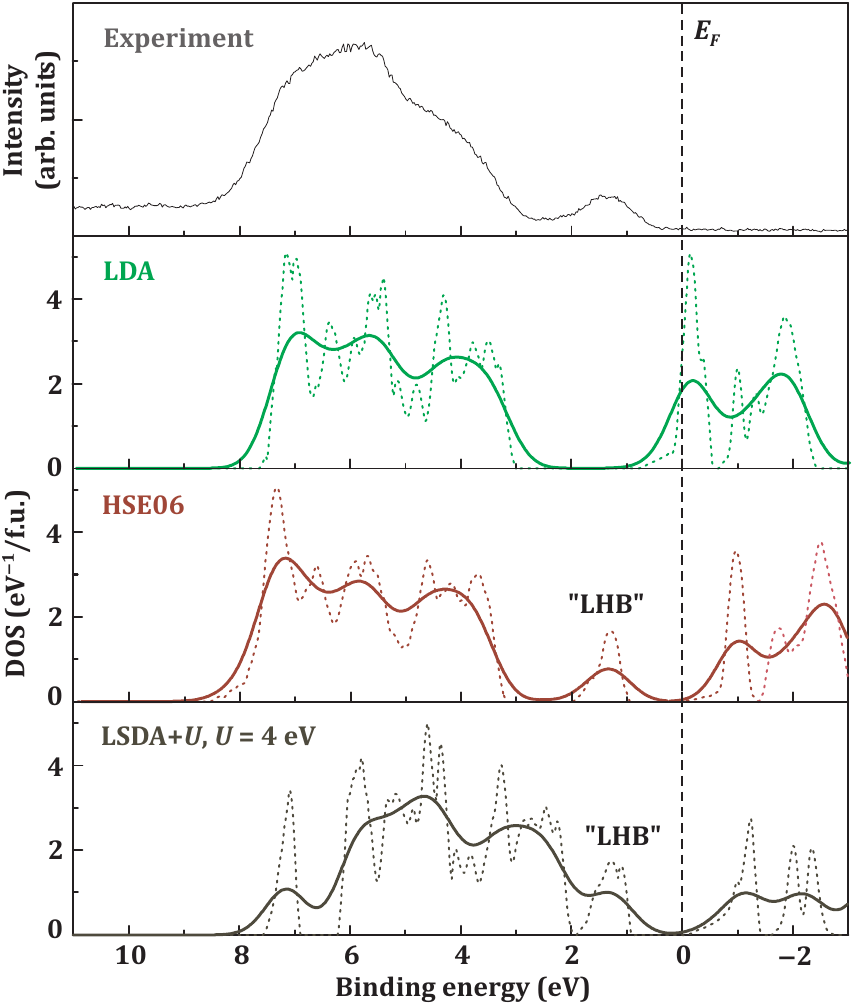}
\caption{\label{fig:spectra}
(Color online) Top panel: valence-band XPS spectrum of YCrO$_4$. Second to bottom panels: density of states (DOS) calculated using LDA, HSE06 hybrid functional, and LSDA+$U$ ($U=4$~eV), respectively. Raw DOS curves (dotted lines) are broadened with the experimental resolution (solid lines).
}
\end{figure}

To determine quantitatively the insulating state of YCrO$_4$, we have carried out an x-ray photoelectron spectroscopy (XPS) experiment. The spectrum was recorded at room temperature in a spectrometer equipped with a Scienta SES-3000 electron energy analyzer and a Vacuum Generators twin crystal monochromatized Al-$K\alpha$ ($h\nu=1486.6$~eV) source. The overall energy resolution was set to 0.4 eV, as determined using the Fermi cutoff of a metallic silver reference, which was also taken as the zero of the binding energy scale. The base pressure in the spectrometer was $2\times 10^{-10}$~mbar and the sample was cleaved \textit{in-situ} to obtain a clean surface. Polycrystalline sample of YCrO$_4$ was prepared by solid-state synthesis as described elsewhere.\cite{mahmoud} The valence-band (VB) spectrum is displayed in the top panel of Fig.~\ref{fig:spectra}. No charging effects were present, which is important to ensure a reliable energy definition.

The top of the valence band is about 0.5~eV away from the Fermi level, thus establishing not only that YCrO$_4$ is a robust insulator, but also that its band gap is at least 0.5~eV. Our VB XPS spectrum is similar but not identical to that of a LaCrO$_4$ thin film.\cite{qiao2013} In the latter, for example, the spectrum is shifted 0.5~eV toward lower binding energies, and the top of the valence band is at the Fermi level. We also note that our spectrum reveals clearly the insulating state, while in an earlier XPS study~\cite{konno1992} the poorer energy resolution of 1.2~eV does not allow one to decide unambiguously on the presence or absence of spectral weight at the Fermi level.

In order to analyze individual spectral features, we first performed band-structure calculations using the full-potential \texttt{FPLO} code with the basis set of local atomic-like orbitals.\cite{fplo} The resulting energy spectrum calculated within the local density approximation (LDA)\cite{pw92} is depicted in the second-from-top panel of Fig.~\ref{fig:spectra} and shows the anticipated splitting of Cr $3d$ states into the crystal-field levels of the $t_2$ and $e$ symmetry (compare to the right panel of Fig.~\ref{fig:str}). Additional splittings related to a weak distortion of the CrO$_4$ tetrahedra are smaller than the bandwidth.

\begin{figure}
\includegraphics{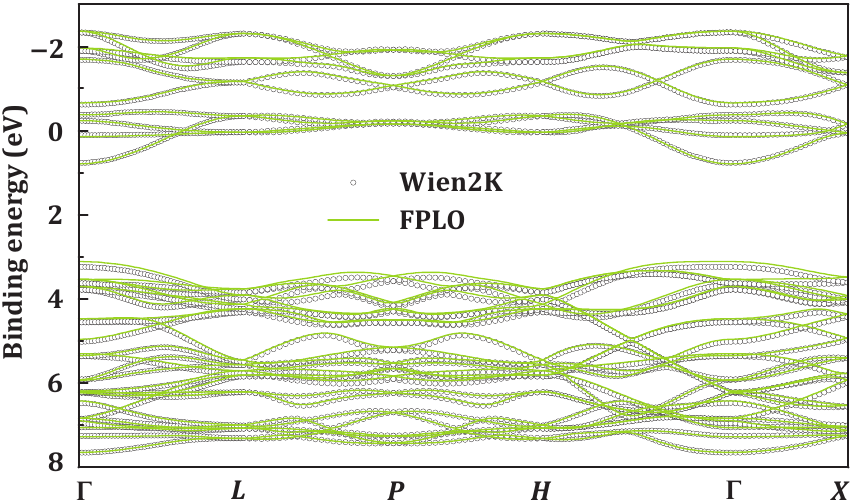}
\caption{\label{fig:comparison}
(Color online) Comparison of LDA band structures calculated in \texttt{FPLO} (lines) and \texttt{Wien2K} (circles) showing that the choice of the basis set has no influence on the accuracy at the LDA level.}
\end{figure}

We can directly conclude that on the LDA level the insulating nature of YCrO$_4$ is not reproduced. The energy splitting between the lowest-energy peak and the broad band at $3-8$~eV binding energy is, correspondingly, substantially larger in the calculation. Suspecting that electron correlation effects may play an important role, we took advantage of the Heyd-Scuseria-Ernzerhof (HSE) hybrid-functional approach,\cite{hse03,*hse04} where a 25\% admixture of the exact (Hartree-Fock) exchange to conventional DFT functionals is incorporated. This calculation has been performed in the \texttt{VASP} code\cite{vasp1,*vasp2}. The outcome is extremely encouraging (Fig.~\ref{fig:spectra}, bottom): the theoretical spectral line shape and energy positions of all features match very well with that of the experiment. The insulating state is also convincingly reproduced with a calculated band gap of 1.0~eV.

\begin{figure}
\includegraphics{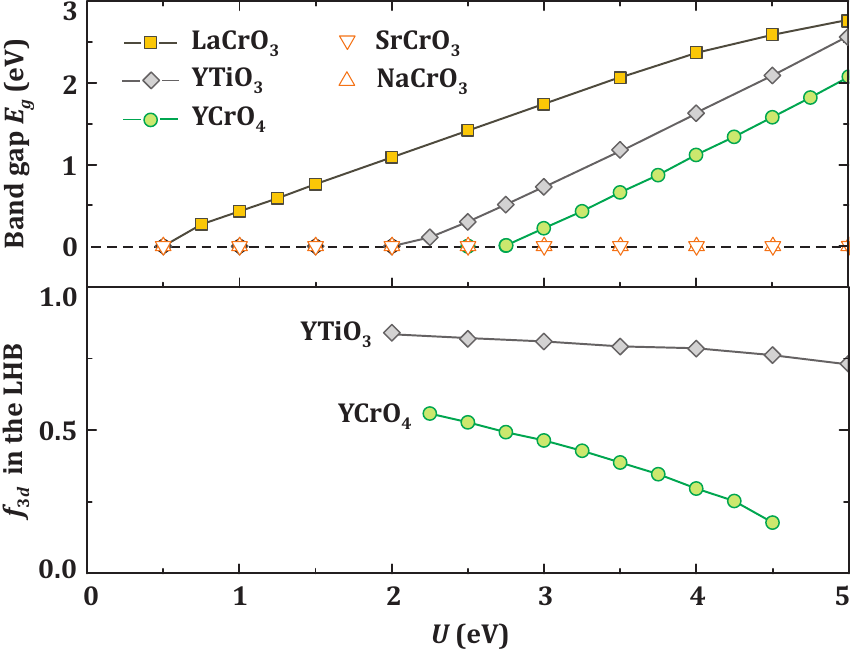}
\caption{\label{fig:u}
(Color online) Top panel: LSDA+$U$ band gap $E_g$ as a function of the on-site Coulomb repulsion $U$ in YCrO$_4$ and several reference compounds ranging from the archetype Mott-Hubbard insulator YTiO$_3$ to negative-$\Delta$ metals SrCrO$_3$ and NaCrO$_3$. Bottom panel: fraction of transition-metal $3d$ states ($f_{3d}$) in the highest occupied state denoted LHB (see Fig.~\ref{fig:spectra}). The fractions are calculated by integrating the atomic-resolved DOS. Above $U=4.5$~eV, the LHB of YCrO$_4$ merges into the oxygen bands, and $f_{3d}$ can not be determined unambiguously.
}
\end{figure}
A similar band gap can be obtained in the LSDA+$U$ approach with a properly adjusted $U$ parameter that accounts for Coulomb correlation in the transition-metal $d$-shell on a mean-field level. However, the separation between the narrow band at $1-2$~eV and the broad band above 3~eV binding energy is not well reproduced (Fig.~\ref{fig:spectra}), because LSDA+$U$ does not correct for the wrong position of oxygen $2p$ levels, which is a well-known problem for a multitude of oxides.\cite{[{For example: }][{}]uddin2006} This problem is remedied by the hybrid functionals. We note in passing that a similar spectrum of YCrO$_4$ has been reported by Li~\textit{et al.}\cite{li2006} who also used the LSDA+$U$ method but with a much higher $U$ value of 7~eV compared to 4~eV in our calculation. As Li~\textit{et al.}\cite{li2006} used a different band-structure code (\texttt{Wien2K}),\cite{wien2k} we also performed an LDA calculation in this code and verified that both codes (\texttt{FPLO} and \texttt{Wien2K}) yield nearly indistinguishable results on the LDA level (Fig.~\ref{fig:comparison}). Therefore, we attribute the difference in the value of $U$ to the different basis sets, which then implies that same $U$ has different effect on the orbital occupations depending on the band-structure code and, in particular, on the basis set.

Remarkably, the DOS for Sr$_3$Cr$_2$O$_8$ calculated by Radtke~\textit{et al.}\cite{radtke2010} is, in general, rather similar to that of YCrO$_4$, although the two compounds have quite different crystal structures. We infer a common origin for this, namely, the presence of Cr$^{5+}$ in the tetrahedral oxygen coordination. It has also been pointed out recently that oxide thin films containing Cr ions in the octahedral coordination have very different properties than those having Cr in the tetrahedral coordination.\cite{qiao2013} In the following, we will focus on the energy characteristics of the Cr states with the goal to understand how the Cr$^{5+}$ systems can be insulators, and to elucidate why the tetrahedral coordination is a necessity for the stabilization of crystal structures containing high Cr valences.

\begin{figure*}
\includegraphics{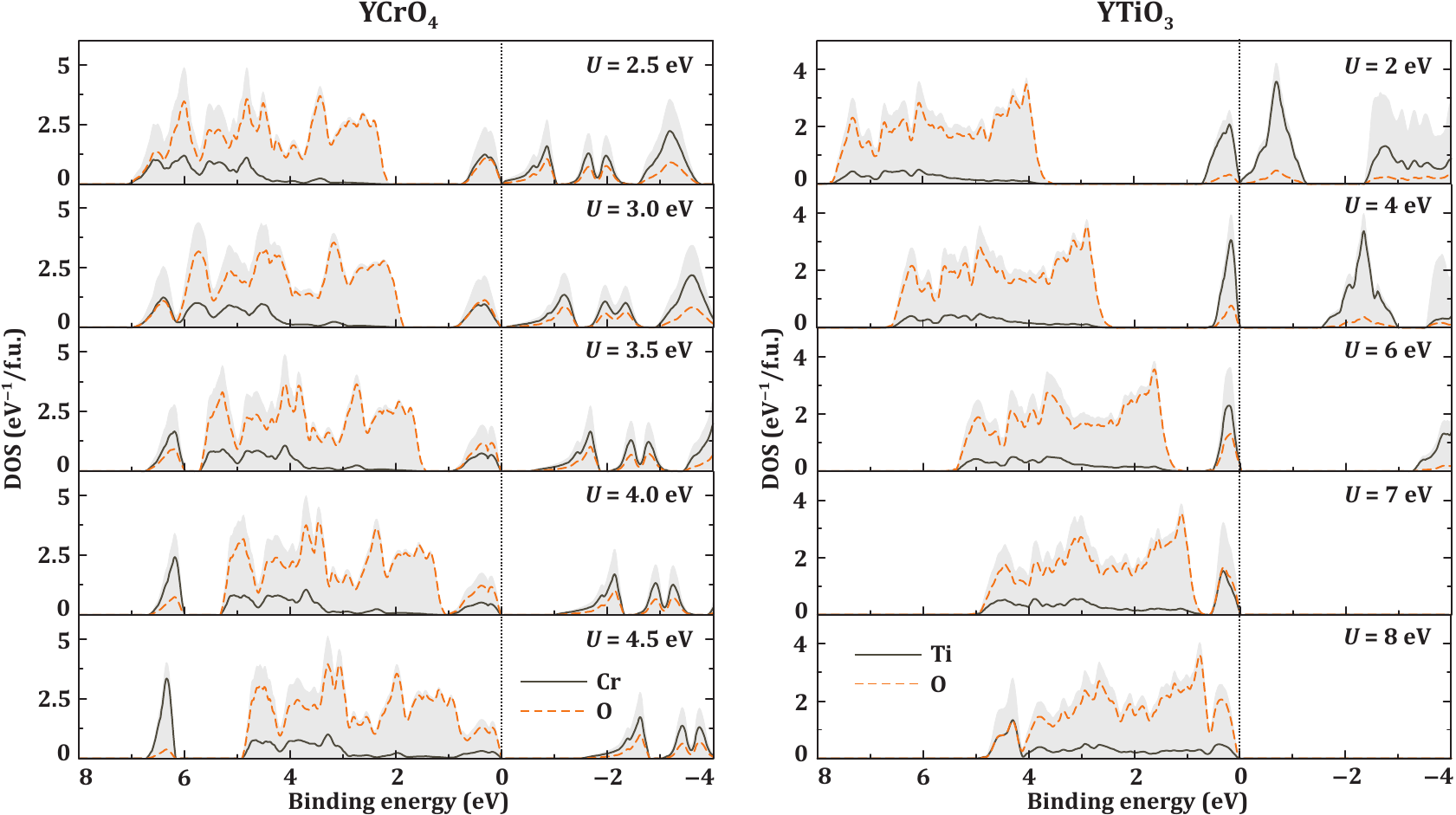}
\caption{\label{fig:ueffect}
(Color online) LSDA+$U$ density of states (DOS) calculated for YCrO$_4$ (left panel) and YTiO$_3$ (right panel) for different values of $U$; only the spin-up channel is shown. Solid and dashed lines are the contributions of the transition-metal (Cr$^{5+}$, Ti$^{3+}$) and ligand (O$^{2-}$) states, respectively. The shading shows the total DOS. The Fermi level is at zero energy and marked by the dotted line that indicates the evolution of YTiO$_3$ from a metal (small $U$) toward a Mott insulator with a well-defined lower Hubbard band (moderate $U$), and eventually toward a charge-transfer insulator with the highest occupied states formed by oxygen (large $U$). In YCrO$_4$, this conventional scenario is altered, because the ``lower Hubbard band'' of this compound features a large contribution of oxygen even at a low $U$ of $2.5-3.0$\,eV, where the band gap is just opened. }
\end{figure*}

To place YCrO$_4$ in the ZSA phase diagram,\cite{zaanen1985} we have carried out band-structure calculations, in which we include specifically a parameter $U$ to account for correlation effects in a mean-field manner. Using this so-called LSDA+$U$ approach,\cite{anisimov1991,anisimov1993} we then investigate how the band gap is affected by a varying magnitude of $U$. The on-site Hund's exchange parameter is set at its standard value of $J=1$~eV.\cite{streltsov2008,lee2009,chioncel2007} The results are summarized in Fig.~\ref{fig:u}, and representative energy spectra are shown in Fig.~\ref{fig:ueffect}. The calculations were performed for YCrO$_4$ together with LaCrO$_3$, SrCrO$_3$, and NaCrO$_3$ as reference systems for different oxidation states of Cr, and for YTiO$_3$ as a reference for a typical Mott-Hubbard insulator. All the calculations assumed ferromagnetic phases, and a simple cubic structure of Cr-based perovskite compounds was chosen in order to simplify the comparison.

Starting with YTiO$_3$ (Fig.~\ref{fig:ueffect}, right), we can clearly see that a band gap is opened for $U$ larger than 2.0~eV, and this band gap increases linearly with $U$ (Fig.~\ref{fig:u}). As expected for a typical Mott-Hubbard insulator, a threshold value for $U$ has to be exceeded in order to overcome the one-electron bandwidth $W$ of about 1.9~eV before the insulating state sets in. LaCrO$_3$ has also a band gap that increases linearly with $U$, and can, therefore, be classified as a Mott-Hubbard insulator as well (Fig.~\ref{fig:u}). Yet, the threshold value for $U$ is very small (about 0.5~eV) despite a sizable $t_{2g}$ bandwidth of about 2.5~eV, and this can be ascribed to the fact that the gap in this $3d^3$ system is between the Cr $t_{2g}$ and $e_g$ bands, so that the gap is facilitated by the octahedral crystal-field splitting. For SrCrO$_3$, we notice that it does not become insulating for a wide range of $U$ values. A similar finding has been obtained for CrO$_2$ by Korotin~\textit{et al.},\cite{korotin1998} who argued that the lack of the band gap in LSDA+$U$ is caused by oxygen $2p$ bands crossing the Fermi level. In CrO$_2$, the presence of oxygen holes associated with the high oxidation state of Cr in this Cr$^{4+}$ system stabilizes the metallic state. This explanation can also be applied to the metallic regime of NaCrO$_3$ persisting even at very large values of $U$, since the Cr valence here is extremely high, namely 5+.

Interestingly, the Cr$^{5+}$ system YCrO$_4$ shows a behavior that is in between YTiO$_3$/LaCrO$_3$ on one hand, and SrCrO$_3$/NaCrO$_3$ on the other hand. YCrO$_4$ does become an insulator for large values of $U$, but only so with a fairly large threshold value of $U=2.75$~eV (see Fig.~\ref{fig:ueffect}, left and Fig.~\ref{fig:u}). This value is much larger than the one-electron band width of the Cr $e$ bands, which is about 1.3~eV, as can be seen from the LDA panel of Fig.~\ref{fig:spectra}. This behavior clearly does not fit to that of a classical Mott-Hubbard insulator, despite the fact that the peak at 1.4~eV binding energy seemingly has all the characteristics of being the lower Hubbard band (``LHB'', see Fig.~\ref{fig:spectra}).

\begin{figure}
\includegraphics{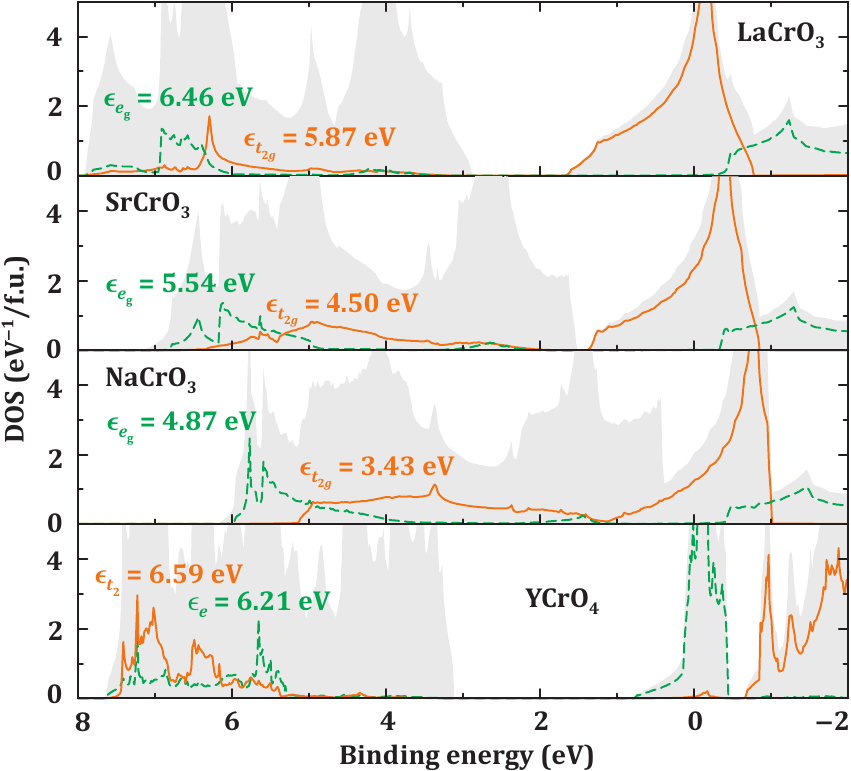}
\caption{\label{fig:dos}
(Color online) Total (shaded) and Cr orbital-resolved DOS for YCrO$_4$ and reference compounds LaCrO$_3$, SrCrO$_3$, and NaCrO$_3$. The Fermi level is at zero energy. The numbers $\epsilon_i$ are relevant centers of gravity for the Cr states hybridized with oxygen.
}
\end{figure}
We now look at the orbital character of this ``LHB'' as a function of $U$ (Fig.~\ref{fig:ueffect}), and make a comparison with the YTiO$_3$ case as a reference. The bottom panel of Fig.~\ref{fig:u} reveals that for the relevant $U=4-5$\,eV\cite{pavarini2004} the LHB of YTiO$_3$ has predominantly $3d$ character and less than one quarter of the O $2p$ states, thus justifying again the description of YTiO$_3$ as a Mott-Hubbard insulator. For YCrO$_4$, however, we observe a quite different behavior. For low values of $U$, where the band gap is just opened, the first ionization states are built up of both Cr $3d$ and O $2p$ orbitals in about equal amounts. Upon increasing $U$, the Cr $3d$ character is reduced and for high values of $U$, the O $2p$ nature prevails, even though this spectral feature remains separated from the bulk of the O $2p$ states. We can, therefore, state that YCrO$_4$ is not a Mott-Hubbard insulator. With a band gap of about 1 eV and a corresponding $U\approx 4$~eV (compare the top panel of Fig.~\ref{fig:u} with the experimental band gap of about 1\,eV), we would rather classify YCrO$_4$ as a charge-transfer insulator. Consequently, this is also the reason why the opening of the band gap requires a high threshold value of $U$. Due to the strong hybridization,\footnote{The hopping integral $V(e)$ between a Cr orbital of $e$ symmetry and its O $2p$ ligands in a CrO$_4$ tetrahedral cluster is about $3.05-3.30$~eV according to LDA [$(pd\pi)=1.87-2.02$~eV], while the hopping integral $V(t_{2g})$ between a $t_{2g}$ orbital and its oxygen ligands in a CrO$_6$ octahedral cluster is substantially smaller, about 2.20~eV for LaCrO$_3$ [$(pd\pi)=1.10$~eV].} the O $2p$ states are able to push firmly against the ``occupied'' Cr $e$ band. Therefore, the energy lowering of this ``occupied'' Cr $e$ band due to $U$ is counteracted, thereby delaying the band gap formation.

Having established the basic characteristics of the electronic structure of YCrO$_4$ in terms of the ZSA diagram, we can now explain why this system is a robust insulator, while the isovalent NaCrO$_3$ and even the lower-valent SrCrO$_3$ and CrO$_2$ can be classified as negative-charge-transfer $p$-type metals. Fig.~\ref{fig:dos} shows the partial density of states of LaCrO$_3$, SrCrO$_3$, NaCrO$_3$ and YCrO$_4$, calculated in the non-magnetic phase using standard LDA. The 113 compounds all have in common that the Cr $3d$ $e_g$ states hybridize with the bottom half of the O $2p$ band and that the Cr $3d$ $t_{2g}$ states tend to hybridize with the top half of the O $2p$ band, especially in NaCrO$_3$. Partially filled states at the Fermi level have the $t_{2g}$ symmetry. Then the relevant action for the band gap formation is determined by the positions of those $t_{2g}$ orbitals with respect to the O $2p$ bands with the corresponding $t_{2g}$ symmetry. The relevant energy scale is, therefore, the energy difference between Cr $t_{2g}$ and the \textit{top} of the O $2p$ band.

Looking now at the YCrO$_4$ case, we can clearly observe that both the Cr $3d$ $t_2$ and $e$ bands mix with the \textit{bottom} of the oxygen band. The relevant charge-transfer energy here is then determined by the Cr $e$ states and the \textit{bottom} of the O $2p$ band. For the 113 compounds, the \textit{effective} charge-transfer energy is reduced by about half of the oxygen band width with respect to the ZSA $\Delta$, which is defined relative to the center of the oxygen band. By contrast, the \textit{effective} charge-transfer energy for YCrO$_4$ is increased by about half of the O $2p$ bandwidth. The difference between the 113 compounds and YCrO$_4$ is obviously caused by the difference in the local coordination of the Cr ions. In the 113 compounds, the Cr ions are octahedrally coordinated by oxygens, while in YCrO$_4$ the coordination of Cr is tetrahedral.

The implications may be generalized for transition-metal oxides with only partially filled $3d$ shells: high oxidation state compounds -- those with the octahedral coordination of the transition metal -- have the potential to feature negative charge-transfer energies and, thus, become metallic, while those with the tetrahedral coordination can still have positive charge-transfer energies and maintain an insulating behavior. Equivalently, one can also state that very high oxidation state materials may have the tendency to build crystal structures with tetrahedral coordination, since the associated increase in the charge-transfer energy ensures the formation of stable insulating states rather than unstable metallic phases with high amount of oxygen holes. 

A further intriguing question is the character of charge carriers induced in such compounds by doping. When transition-metal ions have a large effective $U$, the first ionization state should be positioned at the top of the oxygen band, which is non-bonding. This then leads to the scenario that Korotin \textit{et~al.}\cite{korotin1998} speculated about in their search for \textit{self-doped} charge carriers in CrO$_2$. One can then ask how such a particle will interact with the transition-metal ions. Will it form a new quasi-particle with a spin-quantum number, which is very different from that of transition-metal ions, such that a spin-blockade situation occurs~\cite{maignan2004,chang2009} rendering a high effective mass of this particle? We believe that compounds with the tetrahedral coordination of transition-metal ions deserve further exploration.


\acknowledgments
We would like to acknowledge fruitful discussions with G.A. Sawatzky, D.I. Khomskii, N. Hollmann, and M.W. Haverkort. AT was partly supported by the Mobilitas program of the ESF (grant no. MTT77). The research in Germany is partially supported by the Deutsche Forschungsgemeinschaft through FOR1346 and the work in Spain by the Marie Curie actions -- FP7 -- through SOPRANO ITN (Grant agreement no. 214040).


\begin{thebibliography}{40}%
\makeatletter
\providecommand \@ifxundefined [1]{%
 \@ifx{#1\undefined}
}%
\providecommand \@ifnum [1]{%
 \ifnum #1\expandafter \@firstoftwo
 \else \expandafter \@secondoftwo
 \fi
}%
\providecommand \@ifx [1]{%
 \ifx #1\expandafter \@firstoftwo
 \else \expandafter \@secondoftwo
 \fi
}%
\providecommand \natexlab [1]{#1}%
\providecommand \enquote  [1]{``#1''}%
\providecommand \bibnamefont  [1]{#1}%
\providecommand \bibfnamefont [1]{#1}%
\providecommand \citenamefont [1]{#1}%
\providecommand \href@noop [0]{\@secondoftwo}%
\providecommand \href [0]{\begingroup \@sanitize@url \@href}%
\providecommand \@href[1]{\@@startlink{#1}\@@href}%
\providecommand \@@href[1]{\endgroup#1\@@endlink}%
\providecommand \@sanitize@url [0]{\catcode `\\12\catcode `\$12\catcode
  `\&12\catcode `\#12\catcode `\^12\catcode `\_12\catcode `\%12\relax}%
\providecommand \@@startlink[1]{}%
\providecommand \@@endlink[0]{}%
\providecommand \url  [0]{\begingroup\@sanitize@url \@url }%
\providecommand \@url [1]{\endgroup\@href {#1}{\urlprefix }}%
\providecommand \urlprefix  [0]{URL }%
\providecommand \Eprint [0]{\href }%
\providecommand \doibase [0]{http://dx.doi.org/}%
\providecommand \selectlanguage [0]{\@gobble}%
\providecommand \bibinfo  [0]{\@secondoftwo}%
\providecommand \bibfield  [0]{\@secondoftwo}%
\providecommand \translation [1]{[#1]}%
\providecommand \BibitemOpen [0]{}%
\providecommand \bibitemStop [0]{}%
\providecommand \bibitemNoStop [0]{.\EOS\space}%
\providecommand \EOS [0]{\spacefactor3000\relax}%
\providecommand \BibitemShut  [1]{\csname bibitem#1\endcsname}%
\let\auto@bib@innerbib\@empty
\bibitem [{\citenamefont {Zaanen}\ \emph {et~al.}(1985)\citenamefont {Zaanen},
  \citenamefont {Sawatzky},\ and\ \citenamefont {Allen}}]{zaanen1985}%
  \BibitemOpen
  \bibfield  {author} {\bibinfo {author} {\bibfnamefont {J.}~\bibnamefont
  {Zaanen}}, \bibinfo {author} {\bibfnamefont {G.~A.}\ \bibnamefont
  {Sawatzky}}, \ and\ \bibinfo {author} {\bibfnamefont {J.~W.}\ \bibnamefont
  {Allen}},\ }\href@noop {} {\bibfield  {journal} {\bibinfo  {journal} {Phys.
  Rev. Lett.}\ }\textbf {\bibinfo {volume} {55}},\ \bibinfo {pages} {418}
  (\bibinfo {year} {1985})}\BibitemShut {NoStop}%
\bibitem [{\citenamefont {Korotin}\ \emph {et~al.}(1998)\citenamefont
  {Korotin}, \citenamefont {Anisimov}, \citenamefont {Khomskii},\ and\
  \citenamefont {Sawatzky}}]{korotin1998}%
  \BibitemOpen
  \bibfield  {author} {\bibinfo {author} {\bibfnamefont {M.~A.}\ \bibnamefont
  {Korotin}}, \bibinfo {author} {\bibfnamefont {V.~I.}\ \bibnamefont
  {Anisimov}}, \bibinfo {author} {\bibfnamefont {D.~I.}\ \bibnamefont
  {Khomskii}}, \ and\ \bibinfo {author} {\bibfnamefont {G.~A.}\ \bibnamefont
  {Sawatzky}},\ }\href@noop {} {\bibfield  {journal} {\bibinfo  {journal}
  {Phys. Rev. Lett.}\ }\textbf {\bibinfo {volume} {80}},\ \bibinfo {pages}
  {4305} (\bibinfo {year} {1998})}\BibitemShut {NoStop}%
\bibitem [{\citenamefont {Stagarescu}\ \emph {et~al.}(2000)\citenamefont
  {Stagarescu}, \citenamefont {Su}, \citenamefont {Eastman}, \citenamefont
  {Altmann}, \citenamefont {Himpsel},\ and\ \citenamefont
  {Gupta}}]{stagarescu2000}%
  \BibitemOpen
  \bibfield  {author} {\bibinfo {author} {\bibfnamefont {C.~B.}\ \bibnamefont
  {Stagarescu}}, \bibinfo {author} {\bibfnamefont {X.}~\bibnamefont {Su}},
  \bibinfo {author} {\bibfnamefont {D.~E.}\ \bibnamefont {Eastman}}, \bibinfo
  {author} {\bibfnamefont {K.~N.}\ \bibnamefont {Altmann}}, \bibinfo {author}
  {\bibfnamefont {F.~J.}\ \bibnamefont {Himpsel}}, \ and\ \bibinfo {author}
  {\bibfnamefont {A.}~\bibnamefont {Gupta}},\ }\href@noop {} {\bibfield
  {journal} {\bibinfo  {journal} {Phys. Rev. B}\ }\textbf {\bibinfo {volume}
  {61}},\ \bibinfo {pages} {R9233} (\bibinfo {year} {2000})}\BibitemShut
  {NoStop}%
\bibitem [{\citenamefont {Huang}\ \emph {et~al.}(2003)\citenamefont {Huang},
  \citenamefont {Tjeng}, \citenamefont {Chen}, \citenamefont {Chang},
  \citenamefont {Wu}, \citenamefont {Chung}, \citenamefont {Tanaka},
  \citenamefont {Guo}, \citenamefont {Lin}, \citenamefont {Shyu}, \citenamefont
  {Wu},\ and\ \citenamefont {Chen}}]{huang2003}%
  \BibitemOpen
  \bibfield  {author} {\bibinfo {author} {\bibfnamefont {D.~J.}\ \bibnamefont
  {Huang}}, \bibinfo {author} {\bibfnamefont {L.~H.}\ \bibnamefont {Tjeng}},
  \bibinfo {author} {\bibfnamefont {J.}~\bibnamefont {Chen}}, \bibinfo {author}
  {\bibfnamefont {C.~F.}\ \bibnamefont {Chang}}, \bibinfo {author}
  {\bibfnamefont {W.~P.}\ \bibnamefont {Wu}}, \bibinfo {author} {\bibfnamefont
  {S.~C.}\ \bibnamefont {Chung}}, \bibinfo {author} {\bibfnamefont
  {A.}~\bibnamefont {Tanaka}}, \bibinfo {author} {\bibfnamefont {G.~Y.}\
  \bibnamefont {Guo}}, \bibinfo {author} {\bibfnamefont {H.-J.}\ \bibnamefont
  {Lin}}, \bibinfo {author} {\bibfnamefont {S.~G.}\ \bibnamefont {Shyu}},
  \bibinfo {author} {\bibfnamefont {C.~C.}\ \bibnamefont {Wu}}, \ and\ \bibinfo
  {author} {\bibfnamefont {C.~T.}\ \bibnamefont {Chen}},\ }\href@noop {}
  {\bibfield  {journal} {\bibinfo  {journal} {Phys. Rev. B}\ }\textbf {\bibinfo
  {volume} {67}},\ \bibinfo {pages} {214419} (\bibinfo {year}
  {2003})}\BibitemShut {NoStop}%
\bibitem [{\citenamefont {{Ortega-San-Martin}}\ \emph
  {et~al.}(2007)\citenamefont {{Ortega-San-Martin}}, \citenamefont {Williams},
  \citenamefont {Rodgers}, \citenamefont {Attfield}, \citenamefont {Heymann},\
  and\ \citenamefont {Huppertz}}]{ortega2007}%
  \BibitemOpen
  \bibfield  {author} {\bibinfo {author} {\bibfnamefont {L.}~\bibnamefont
  {{Ortega-San-Martin}}}, \bibinfo {author} {\bibfnamefont {A.}~\bibnamefont
  {Williams}}, \bibinfo {author} {\bibfnamefont {J.}~\bibnamefont {Rodgers}},
  \bibinfo {author} {\bibfnamefont {J.~P.}\ \bibnamefont {Attfield}}, \bibinfo
  {author} {\bibfnamefont {G.}~\bibnamefont {Heymann}}, \ and\ \bibinfo
  {author} {\bibfnamefont {H.}~\bibnamefont {Huppertz}},\ }\href@noop {}
  {\bibfield  {journal} {\bibinfo  {journal} {Phys. Rev. Lett.}\ }\textbf
  {\bibinfo {volume} {99}},\ \bibinfo {pages} {255701} (\bibinfo {year}
  {2007})}\BibitemShut {NoStop}%
\bibitem [{\citenamefont {Lee}\ and\ \citenamefont {Pickett}(2009)}]{lee2009}%
  \BibitemOpen
  \bibfield  {author} {\bibinfo {author} {\bibfnamefont {K.-W.}\ \bibnamefont
  {Lee}}\ and\ \bibinfo {author} {\bibfnamefont {W.~E.}\ \bibnamefont
  {Pickett}},\ }\href@noop {} {\bibfield  {journal} {\bibinfo  {journal} {Phys.
  Rev. B}\ }\textbf {\bibinfo {volume} {80}},\ \bibinfo {pages} {125133}
  (\bibinfo {year} {2009})}\BibitemShut {NoStop}%
\bibitem [{\citenamefont {Komarek}\ \emph {et~al.}(2008)\citenamefont
  {Komarek}, \citenamefont {Streltsov}, \citenamefont {Isobe}, \citenamefont
  {M\"oller}, \citenamefont {Hoelzel}, \citenamefont {Senyshyn}, \citenamefont
  {Trots}, \citenamefont {Fern{\'a}ndez-D{\i}az}, \citenamefont {Hansen},
  \citenamefont {Gotou}, \citenamefont {Yagi}, \citenamefont {Ueda},
  \citenamefont {Anisimov}, \citenamefont {Gr\"uninger}, \citenamefont
  {Khomskii},\ and\ \citenamefont {Braden}}]{komarek2008}%
  \BibitemOpen
  \bibfield  {author} {\bibinfo {author} {\bibfnamefont {A.~C.}\ \bibnamefont
  {Komarek}}, \bibinfo {author} {\bibfnamefont {S.~V.}\ \bibnamefont
  {Streltsov}}, \bibinfo {author} {\bibfnamefont {M.}~\bibnamefont {Isobe}},
  \bibinfo {author} {\bibfnamefont {T.}~\bibnamefont {M\"oller}}, \bibinfo
  {author} {\bibfnamefont {M.}~\bibnamefont {Hoelzel}}, \bibinfo {author}
  {\bibfnamefont {A.}~\bibnamefont {Senyshyn}}, \bibinfo {author}
  {\bibfnamefont {D.}~\bibnamefont {Trots}}, \bibinfo {author} {\bibfnamefont
  {M.~T.}\ \bibnamefont {Fern{\'a}ndez-D{\i}az}}, \bibinfo {author}
  {\bibfnamefont {T.}~\bibnamefont {Hansen}}, \bibinfo {author} {\bibfnamefont
  {H.}~\bibnamefont {Gotou}}, \bibinfo {author} {\bibfnamefont
  {T.}~\bibnamefont {Yagi}}, \bibinfo {author} {\bibfnamefont {Y.}~\bibnamefont
  {Ueda}}, \bibinfo {author} {\bibfnamefont {V.~I.}\ \bibnamefont {Anisimov}},
  \bibinfo {author} {\bibfnamefont {M.}~\bibnamefont {Gr\"uninger}}, \bibinfo
  {author} {\bibfnamefont {D.~I.}\ \bibnamefont {Khomskii}}, \ and\ \bibinfo
  {author} {\bibfnamefont {M.}~\bibnamefont {Braden}},\ }\href@noop {}
  {\bibfield  {journal} {\bibinfo  {journal} {Phys. Rev. Lett.}\ }\textbf
  {\bibinfo {volume} {101}},\ \bibinfo {pages} {167204} (\bibinfo {year}
  {2008})}\BibitemShut {NoStop}%
\bibitem [{\citenamefont {Streltsov}\ \emph {et~al.}(2008)\citenamefont
  {Streltsov}, \citenamefont {Korotin}, \citenamefont {Anisimov},\ and\
  \citenamefont {Khomskii}}]{streltsov2008}%
  \BibitemOpen
  \bibfield  {author} {\bibinfo {author} {\bibfnamefont {S.~V.}\ \bibnamefont
  {Streltsov}}, \bibinfo {author} {\bibfnamefont {M.~A.}\ \bibnamefont
  {Korotin}}, \bibinfo {author} {\bibfnamefont {V.~I.}\ \bibnamefont
  {Anisimov}}, \ and\ \bibinfo {author} {\bibfnamefont {D.~I.}\ \bibnamefont
  {Khomskii}},\ }\href@noop {} {\bibfield  {journal} {\bibinfo  {journal}
  {Phys. Rev. B}\ }\textbf {\bibinfo {volume} {78}},\ \bibinfo {pages} {054425}
  (\bibinfo {year} {2008})}\BibitemShut {NoStop}%
\bibitem [{\citenamefont {Komarek}\ \emph {et~al.}(2011)\citenamefont
  {Komarek}, \citenamefont {M\"oller}, \citenamefont {Isobe}, \citenamefont
  {Drees}, \citenamefont {Ulbrich}, \citenamefont {Azuma}, \citenamefont
  {Fern\'andez-D{\'\i}az}, \citenamefont {Senyshyn}, \citenamefont {Hoelzel},
  \citenamefont {Andr\'e}, \citenamefont {Ueda}, \citenamefont {Gr\"uninger},\
  and\ \citenamefont {Braden}}]{komarek2011}%
  \BibitemOpen
  \bibfield  {author} {\bibinfo {author} {\bibfnamefont {A.~C.}\ \bibnamefont
  {Komarek}}, \bibinfo {author} {\bibfnamefont {T.}~\bibnamefont {M\"oller}},
  \bibinfo {author} {\bibfnamefont {M.}~\bibnamefont {Isobe}}, \bibinfo
  {author} {\bibfnamefont {Y.}~\bibnamefont {Drees}}, \bibinfo {author}
  {\bibfnamefont {H.}~\bibnamefont {Ulbrich}}, \bibinfo {author} {\bibfnamefont
  {M.}~\bibnamefont {Azuma}}, \bibinfo {author} {\bibfnamefont {M.~T.}\
  \bibnamefont {Fern\'andez-D{\'\i}az}}, \bibinfo {author} {\bibfnamefont
  {A.}~\bibnamefont {Senyshyn}}, \bibinfo {author} {\bibfnamefont
  {M.}~\bibnamefont {Hoelzel}}, \bibinfo {author} {\bibfnamefont
  {G.}~\bibnamefont {Andr\'e}}, \bibinfo {author} {\bibfnamefont
  {Y.}~\bibnamefont {Ueda}}, \bibinfo {author} {\bibfnamefont {M.}~\bibnamefont
  {Gr\"uninger}}, \ and\ \bibinfo {author} {\bibfnamefont {M.}~\bibnamefont
  {Braden}},\ }\href@noop {} {\bibfield  {journal} {\bibinfo  {journal} {Phys.
  Rev. B}\ }\textbf {\bibinfo {volume} {84}},\ \bibinfo {pages} {125114}
  (\bibinfo {year} {2011})}\BibitemShut {NoStop}%
\bibitem [{\citenamefont {Jim\'enez}\ \emph {et~al.}(2000)\citenamefont
  {Jim\'enez}, \citenamefont {Isasi},\ and\ \citenamefont
  {S\'aez-Puche}}]{jimenez2000}%
  \BibitemOpen
  \bibfield  {author} {\bibinfo {author} {\bibfnamefont {E.}~\bibnamefont
  {Jim\'enez}}, \bibinfo {author} {\bibfnamefont {J.}~\bibnamefont {Isasi}}, \
  and\ \bibinfo {author} {\bibfnamefont {R.}~\bibnamefont {S\'aez-Puche}},\
  }\href@noop {} {\bibfield  {journal} {\bibinfo  {journal} {J. Alloys Comp.}\
  }\textbf {\bibinfo {volume} {312}},\ \bibinfo {pages} {53} (\bibinfo {year}
  {2000})}\BibitemShut {NoStop}%
\bibitem [{\citenamefont {Aoki}\ \emph {et~al.}(2001)\citenamefont {Aoki},
  \citenamefont {Konno},\ and\ \citenamefont {Tachikawa}}]{aoki2001}%
  \BibitemOpen
  \bibfield  {author} {\bibinfo {author} {\bibfnamefont {Y.}~\bibnamefont
  {Aoki}}, \bibinfo {author} {\bibfnamefont {H.}~\bibnamefont {Konno}}, \ and\
  \bibinfo {author} {\bibfnamefont {H.}~\bibnamefont {Tachikawa}},\ }\href@noop
  {} {\bibfield  {journal} {\bibinfo  {journal} {J. Mater. Chem.}\ }\textbf
  {\bibinfo {volume} {11}},\ \bibinfo {pages} {1214} (\bibinfo {year}
  {2001})}\BibitemShut {NoStop}%
\bibitem [{\citenamefont {S\'aez-Puche}\ \emph {et~al.}(2003)\citenamefont
  {S\'aez-Puche}, \citenamefont {Jim\'enez}, \citenamefont {Isasi},
  \citenamefont {az},\ and\ \citenamefont {a~Mu\~noz}}]{rcro4-2003}%
  \BibitemOpen
  \bibfield  {author} {\bibinfo {author} {\bibfnamefont {R.}~\bibnamefont
  {S\'aez-Puche}}, \bibinfo {author} {\bibfnamefont {E.}~\bibnamefont
  {Jim\'enez}}, \bibinfo {author} {\bibfnamefont {J.}~\bibnamefont {Isasi}},
  \bibinfo {author} {\bibfnamefont {M.~T. F.-D.}\ \bibnamefont {az}}, \ and\
  \bibinfo {author} {\bibfnamefont {J.~L.~G.}\ \bibnamefont {a~Mu\~noz}},\
  }\href@noop {} {\bibfield  {journal} {\bibinfo  {journal} {J. Solid State
  Chem.}\ }\textbf {\bibinfo {volume} {171}},\ \bibinfo {pages} {161} (\bibinfo
  {year} {2003})}\BibitemShut {NoStop}%
\bibitem [{\citenamefont {Long}\ \emph {et~al.}(2007)\citenamefont {Long},
  \citenamefont {Yang}, \citenamefont {Yu}, \citenamefont {Li}, \citenamefont
  {Yu},\ and\ \citenamefont {Jin}}]{long2007}%
  \BibitemOpen
  \bibfield  {author} {\bibinfo {author} {\bibfnamefont {Y.~W.}\ \bibnamefont
  {Long}}, \bibinfo {author} {\bibfnamefont {L.~X.}\ \bibnamefont {Yang}},
  \bibinfo {author} {\bibfnamefont {Y.}~\bibnamefont {Yu}}, \bibinfo {author}
  {\bibfnamefont {F.~Y.}\ \bibnamefont {Li}}, \bibinfo {author} {\bibfnamefont
  {R.~C.}\ \bibnamefont {Yu}}, \ and\ \bibinfo {author} {\bibfnamefont {C.~Q.}\
  \bibnamefont {Jin}},\ }\href@noop {} {\bibfield  {journal} {\bibinfo
  {journal} {Phys. Rev. B}\ }\textbf {\bibinfo {volume} {75}},\ \bibinfo
  {pages} {104402} (\bibinfo {year} {2007})}\BibitemShut {NoStop}%
\bibitem [{\citenamefont {Errandonea}\ \emph {et~al.}(2011)\citenamefont
  {Errandonea}, \citenamefont {Kumar}, \citenamefont {L{\'o}pez-Solano},
  \citenamefont {Rodr{\'\i}guez-Hern{\'a}ndez}, \citenamefont {Mu{\~n}oz},
  \citenamefont {Rabie},\ and\ \citenamefont {Puche}}]{errandonea2011}%
  \BibitemOpen
  \bibfield  {author} {\bibinfo {author} {\bibfnamefont {D.}~\bibnamefont
  {Errandonea}}, \bibinfo {author} {\bibfnamefont {R.}~\bibnamefont {Kumar}},
  \bibinfo {author} {\bibfnamefont {J.}~\bibnamefont {L{\'o}pez-Solano}},
  \bibinfo {author} {\bibfnamefont {P.}~\bibnamefont
  {Rodr{\'\i}guez-Hern{\'a}ndez}}, \bibinfo {author} {\bibfnamefont
  {A.}~\bibnamefont {Mu{\~n}oz}}, \bibinfo {author} {\bibfnamefont {M.~G.}\
  \bibnamefont {Rabie}}, \ and\ \bibinfo {author} {\bibfnamefont {R.~S.}\
  \bibnamefont {Puche}},\ }\href@noop {} {\bibfield  {journal} {\bibinfo
  {journal} {Phys. Rev. B}\ }\textbf {\bibinfo {volume} {83}},\ \bibinfo
  {pages} {134109} (\bibinfo {year} {2011})}\BibitemShut {NoStop}%
\bibitem [{\citenamefont {Qiao}\ \emph {et~al.}(2013)\citenamefont {Qiao},
  \citenamefont {Xiao}, \citenamefont {Heald}, \citenamefont {Bowden},
  \citenamefont {Varga}, \citenamefont {Exarhos}, \citenamefont {Biegalski},
  \citenamefont {Ivanov}, \citenamefont {Weber}, \citenamefont {Droubaya},\
  and\ \citenamefont {Chambers}}]{qiao2013}%
  \BibitemOpen
  \bibfield  {author} {\bibinfo {author} {\bibfnamefont {L.}~\bibnamefont
  {Qiao}}, \bibinfo {author} {\bibfnamefont {H.~Y.}\ \bibnamefont {Xiao}},
  \bibinfo {author} {\bibfnamefont {S.~M.}\ \bibnamefont {Heald}}, \bibinfo
  {author} {\bibfnamefont {M.~E.}\ \bibnamefont {Bowden}}, \bibinfo {author}
  {\bibfnamefont {T.}~\bibnamefont {Varga}}, \bibinfo {author} {\bibfnamefont
  {G.~J.}\ \bibnamefont {Exarhos}}, \bibinfo {author} {\bibfnamefont {M.~D.}\
  \bibnamefont {Biegalski}}, \bibinfo {author} {\bibfnamefont {I.~N.}\
  \bibnamefont {Ivanov}}, \bibinfo {author} {\bibfnamefont {W.~J.}\
  \bibnamefont {Weber}}, \bibinfo {author} {\bibfnamefont {T.~C.}\ \bibnamefont
  {Droubaya}}, \ and\ \bibinfo {author} {\bibfnamefont {S.~A.}\ \bibnamefont
  {Chambers}},\ }\href@noop {} {\bibfield  {journal} {\bibinfo  {journal} {J.
  Mater. Chem. C}\ }\textbf {\bibinfo {volume} {1}},\ \bibinfo {pages} {4527}
  (\bibinfo {year} {2013})}\BibitemShut {NoStop}%
\bibitem [{\citenamefont {Torrance}\ \emph {et~al.}(1991)\citenamefont
  {Torrance}, \citenamefont {Lacorre}, \citenamefont {Asavaroengchai},\ and\
  \citenamefont {Metzger}}]{torrance1991}%
  \BibitemOpen
  \bibfield  {author} {\bibinfo {author} {\bibfnamefont {J.~B.}\ \bibnamefont
  {Torrance}}, \bibinfo {author} {\bibfnamefont {P.}~\bibnamefont {Lacorre}},
  \bibinfo {author} {\bibfnamefont {C.}~\bibnamefont {Asavaroengchai}}, \ and\
  \bibinfo {author} {\bibfnamefont {R.~M.}\ \bibnamefont {Metzger}},\
  }\href@noop {} {\bibfield  {journal} {\bibinfo  {journal} {Physica C}\
  }\textbf {\bibinfo {volume} {182}},\ \bibinfo {pages} {351} (\bibinfo {year}
  {1991})}\BibitemShut {NoStop}%
\bibitem [{\citenamefont {Radtke}\ \emph {et~al.}(2010)\citenamefont {Radtke},
  \citenamefont {Sa\'ul}, \citenamefont {Dabkowska}, \citenamefont {Luke},\
  and\ \citenamefont {Botton}}]{radtke2010}%
  \BibitemOpen
  \bibfield  {author} {\bibinfo {author} {\bibfnamefont {G.}~\bibnamefont
  {Radtke}}, \bibinfo {author} {\bibfnamefont {A.}~\bibnamefont {Sa\'ul}},
  \bibinfo {author} {\bibfnamefont {H.~A.}\ \bibnamefont {Dabkowska}}, \bibinfo
  {author} {\bibfnamefont {G.~M.}\ \bibnamefont {Luke}}, \ and\ \bibinfo
  {author} {\bibfnamefont {G.~A.}\ \bibnamefont {Botton}},\ }\href@noop {}
  {\bibfield  {journal} {\bibinfo  {journal} {Phys. Rev. Lett.}\ }\textbf
  {\bibinfo {volume} {105}},\ \bibinfo {pages} {036401} (\bibinfo {year}
  {2010})}\BibitemShut {NoStop}%
\bibitem [{\citenamefont {Singh}\ and\ \citenamefont
  {Johnston}(2007)}]{singh2007}%
  \BibitemOpen
  \bibfield  {author} {\bibinfo {author} {\bibfnamefont {Y.}~\bibnamefont
  {Singh}}\ and\ \bibinfo {author} {\bibfnamefont {D.~C.}\ \bibnamefont
  {Johnston}},\ }\href@noop {} {\bibfield  {journal} {\bibinfo  {journal}
  {Phys. Rev. B}\ }\textbf {\bibinfo {volume} {76}},\ \bibinfo {pages} {012407}
  (\bibinfo {year} {2007})}\BibitemShut {NoStop}%
\bibitem [{\citenamefont {Aczel}\ \emph {et~al.}(2009)\citenamefont {Aczel},
  \citenamefont {Kohama}, \citenamefont {Marcenat}, \citenamefont {Weickert},
  \citenamefont {Jaime}, \citenamefont {Ayala-Valenzuela}, \citenamefont
  {{McDonald}}, \citenamefont {Selesnic}, \citenamefont {Dabkowska},\ and\
  \citenamefont {Luke}}]{aczel2009}%
  \BibitemOpen
  \bibfield  {author} {\bibinfo {author} {\bibfnamefont {A.~A.}\ \bibnamefont
  {Aczel}}, \bibinfo {author} {\bibfnamefont {Y.}~\bibnamefont {Kohama}},
  \bibinfo {author} {\bibfnamefont {C.}~\bibnamefont {Marcenat}}, \bibinfo
  {author} {\bibfnamefont {F.}~\bibnamefont {Weickert}}, \bibinfo {author}
  {\bibfnamefont {M.}~\bibnamefont {Jaime}}, \bibinfo {author} {\bibfnamefont
  {O.~E.}\ \bibnamefont {Ayala-Valenzuela}}, \bibinfo {author} {\bibfnamefont
  {R.~D.}\ \bibnamefont {{McDonald}}}, \bibinfo {author} {\bibfnamefont
  {S.~D.}\ \bibnamefont {Selesnic}}, \bibinfo {author} {\bibfnamefont {H.~A.}\
  \bibnamefont {Dabkowska}}, \ and\ \bibinfo {author} {\bibfnamefont {G.~M.}\
  \bibnamefont {Luke}},\ }\href@noop {} {\bibfield  {journal} {\bibinfo
  {journal} {Phys. Rev. Lett.}\ }\textbf {\bibinfo {volume} {103}},\ \bibinfo
  {pages} {207203} (\bibinfo {year} {2009})}\BibitemShut {NoStop}%
\bibitem [{\citenamefont {Kofu}\ \emph {et~al.}(2009)\citenamefont {Kofu},
  \citenamefont {Kim}, \citenamefont {Ji}, \citenamefont {Lee}, \citenamefont
  {Ueda}, \citenamefont {Qiu}, \citenamefont {Kang}, \citenamefont {Green},\
  and\ \citenamefont {Ueda}}]{kofu2009}%
  \BibitemOpen
  \bibfield  {author} {\bibinfo {author} {\bibfnamefont {M.}~\bibnamefont
  {Kofu}}, \bibinfo {author} {\bibfnamefont {J.-H.}\ \bibnamefont {Kim}},
  \bibinfo {author} {\bibfnamefont {S.}~\bibnamefont {Ji}}, \bibinfo {author}
  {\bibfnamefont {S.-H.}\ \bibnamefont {Lee}}, \bibinfo {author} {\bibfnamefont
  {H.}~\bibnamefont {Ueda}}, \bibinfo {author} {\bibfnamefont {Y.}~\bibnamefont
  {Qiu}}, \bibinfo {author} {\bibfnamefont {H.-J.}\ \bibnamefont {Kang}},
  \bibinfo {author} {\bibfnamefont {M.~A.}\ \bibnamefont {Green}}, \ and\
  \bibinfo {author} {\bibfnamefont {Y.}~\bibnamefont {Ueda}},\ }\href@noop {}
  {\bibfield  {journal} {\bibinfo  {journal} {Phys. Rev. Lett.}\ }\textbf
  {\bibinfo {volume} {102}},\ \bibinfo {pages} {037206} (\bibinfo {year}
  {2009})}\BibitemShut {NoStop}%
\bibitem [{\citenamefont {Wang}\ \emph {et~al.}(2011)\citenamefont {Wang},
  \citenamefont {Schmidt}, \citenamefont {G\"unther}, \citenamefont {Schaile},
  \citenamefont {Pascher}, \citenamefont {Mayr}, \citenamefont {Goncharov},
  \citenamefont {Quintero-Castro}, \citenamefont {Islam}, \citenamefont {Lake},
  \citenamefont {{Krug von Nidda}}, \citenamefont {Loidl},\ and\ \citenamefont
  {Deisenhofer}}]{wang2011}%
  \BibitemOpen
  \bibfield  {author} {\bibinfo {author} {\bibfnamefont {Z.}~\bibnamefont
  {Wang}}, \bibinfo {author} {\bibfnamefont {M.}~\bibnamefont {Schmidt}},
  \bibinfo {author} {\bibfnamefont {A.}~\bibnamefont {G\"unther}}, \bibinfo
  {author} {\bibfnamefont {S.}~\bibnamefont {Schaile}}, \bibinfo {author}
  {\bibfnamefont {N.}~\bibnamefont {Pascher}}, \bibinfo {author} {\bibfnamefont
  {F.}~\bibnamefont {Mayr}}, \bibinfo {author} {\bibfnamefont {Y.}~\bibnamefont
  {Goncharov}}, \bibinfo {author} {\bibfnamefont {D.~L.}\ \bibnamefont
  {Quintero-Castro}}, \bibinfo {author} {\bibfnamefont {A.~T. M.~N.}\
  \bibnamefont {Islam}}, \bibinfo {author} {\bibfnamefont {B.}~\bibnamefont
  {Lake}}, \bibinfo {author} {\bibfnamefont {H.-A.}\ \bibnamefont {{Krug von
  Nidda}}}, \bibinfo {author} {\bibfnamefont {A.}~\bibnamefont {Loidl}}, \ and\
  \bibinfo {author} {\bibfnamefont {J.}~\bibnamefont {Deisenhofer}},\
  }\href@noop {} {\bibfield  {journal} {\bibinfo  {journal} {Phys. Rev. B}\
  }\textbf {\bibinfo {volume} {83}},\ \bibinfo {pages} {201102(R)} (\bibinfo
  {year} {2011})}\BibitemShut {NoStop}%
\bibitem [{\citenamefont {Kantorovich}()}]{tetr}%
  \BibitemOpen
  \bibfield  {author} {\bibinfo {author} {\bibfnamefont {L.~N.}\ \bibnamefont
  {Kantorovich}},\ }\href@noop {} {}\bibinfo {note} {User-friendly
  visualization program for \textit{ab initio} DFT codes \texttt{VASP},
  \texttt{SIESTA}, \texttt{QE} and \texttt{QUICKSTEP}, 1996--2012,
  unpublished}\BibitemShut {NoStop}%
\bibitem [{\citenamefont {Rabie}()}]{mahmoud}%
  \BibitemOpen
  \bibfield  {author} {\bibinfo {author} {\bibfnamefont {M.}~\bibnamefont
  {Rabie}},\ }\href@noop {} {}\bibinfo {note} {PhD thesis, University of Madrid
  (2012)}\BibitemShut {NoStop}%
\bibitem [{\citenamefont {Konno}\ \emph {et~al.}(1992)\citenamefont {Konno},
  \citenamefont {Tachikawa}, \citenamefont {Furusaki},\ and\ \citenamefont
  {Furuichi}}]{konno1992}%
  \BibitemOpen
  \bibfield  {author} {\bibinfo {author} {\bibfnamefont {H.}~\bibnamefont
  {Konno}}, \bibinfo {author} {\bibfnamefont {H.}~\bibnamefont {Tachikawa}},
  \bibinfo {author} {\bibfnamefont {A.}~\bibnamefont {Furusaki}}, \ and\
  \bibinfo {author} {\bibfnamefont {R.}~\bibnamefont {Furuichi}},\ }\href@noop
  {} {\bibfield  {journal} {\bibinfo  {journal} {Analyt. Sci.}\ }\textbf
  {\bibinfo {volume} {8}},\ \bibinfo {pages} {641} (\bibinfo {year}
  {1992})}\BibitemShut {NoStop}%
\bibitem [{\citenamefont {Koepernik}\ and\ \citenamefont
  {Eschrig}(1999)}]{fplo}%
  \BibitemOpen
  \bibfield  {author} {\bibinfo {author} {\bibfnamefont {K.}~\bibnamefont
  {Koepernik}}\ and\ \bibinfo {author} {\bibfnamefont {H.}~\bibnamefont
  {Eschrig}},\ }\href@noop {} {\bibfield  {journal} {\bibinfo  {journal} {Phys.
  Rev. B}\ }\textbf {\bibinfo {volume} {59}},\ \bibinfo {pages} {1743}
  (\bibinfo {year} {1999})}\BibitemShut {NoStop}%
\bibitem [{\citenamefont {Perdew}\ and\ \citenamefont {Wang}(1992)}]{pw92}%
  \BibitemOpen
  \bibfield  {author} {\bibinfo {author} {\bibfnamefont {J.~P.}\ \bibnamefont
  {Perdew}}\ and\ \bibinfo {author} {\bibfnamefont {Y.}~\bibnamefont {Wang}},\
  }\href@noop {} {\bibfield  {journal} {\bibinfo  {journal} {Phys. Rev. B}\
  }\textbf {\bibinfo {volume} {45}},\ \bibinfo {pages} {13244} (\bibinfo {year}
  {1992})}\BibitemShut {NoStop}%
\bibitem [{\citenamefont {Heyd}\ \emph {et~al.}(2003)\citenamefont {Heyd},
  \citenamefont {Scuseria},\ and\ \citenamefont {Ernzerhof}}]{hse03}%
  \BibitemOpen
  \bibfield  {author} {\bibinfo {author} {\bibfnamefont {J.}~\bibnamefont
  {Heyd}}, \bibinfo {author} {\bibfnamefont {G.~E.}\ \bibnamefont {Scuseria}},
  \ and\ \bibinfo {author} {\bibfnamefont {M.}~\bibnamefont {Ernzerhof}},\
  }\href@noop {} {\bibfield  {journal} {\bibinfo  {journal} {J. Chem. Phys.}\
  }\textbf {\bibinfo {volume} {118}},\ \bibinfo {pages} {8207} (\bibinfo {year}
  {2003})}\BibitemShut {NoStop}%
\bibitem [{\citenamefont {Heyd}\ and\ \citenamefont {Scuseria}(2004)}]{hse04}%
  \BibitemOpen
  \bibfield  {author} {\bibinfo {author} {\bibfnamefont {J.}~\bibnamefont
  {Heyd}}\ and\ \bibinfo {author} {\bibfnamefont {G.~E.}\ \bibnamefont
  {Scuseria}},\ }\href@noop {} {\bibfield  {journal} {\bibinfo  {journal} {J.
  Chem. Phys.}\ }\textbf {\bibinfo {volume} {121}},\ \bibinfo {pages} {1187}
  (\bibinfo {year} {2004})}\BibitemShut {NoStop}%
\bibitem [{\citenamefont {Kresse}\ and\ \citenamefont
  {Furthm\"uller}(1996{\natexlab{a}})}]{vasp1}%
  \BibitemOpen
  \bibfield  {author} {\bibinfo {author} {\bibfnamefont {G.}~\bibnamefont
  {Kresse}}\ and\ \bibinfo {author} {\bibfnamefont {J.}~\bibnamefont
  {Furthm\"uller}},\ }\href@noop {} {\bibfield  {journal} {\bibinfo  {journal}
  {Comput. Mater. Sci.}\ }\textbf {\bibinfo {volume} {6}},\ \bibinfo {pages}
  {15} (\bibinfo {year} {1996}{\natexlab{a}})}\BibitemShut {NoStop}%
\bibitem [{\citenamefont {Kresse}\ and\ \citenamefont
  {Furthm\"uller}(1996{\natexlab{b}})}]{vasp2}%
  \BibitemOpen
  \bibfield  {author} {\bibinfo {author} {\bibfnamefont {G.}~\bibnamefont
  {Kresse}}\ and\ \bibinfo {author} {\bibfnamefont {J.}~\bibnamefont
  {Furthm\"uller}},\ }\href@noop {} {\bibfield  {journal} {\bibinfo  {journal}
  {Phys. Rev. B}\ }\textbf {\bibinfo {volume} {54}},\ \bibinfo {pages} {11169}
  (\bibinfo {year} {1996}{\natexlab{b}})}\BibitemShut {NoStop}%
\bibitem [{\citenamefont {Uddin}\ and\ \citenamefont
  {Scuseria}(2006)}]{uddin2006}%
  \BibitemOpen
  \bibfield  {author} {\bibinfo {author} {\bibfnamefont {J.}~\bibnamefont
  {Uddin}}\ and\ \bibinfo {author} {\bibfnamefont {G.~E.}\ \bibnamefont
  {Scuseria}},\ }\href@noop {} {\bibfield  {journal} {\bibinfo  {journal}
  {Phys. Rev. B}\ }\textbf {\bibinfo {volume} {74}},\ \bibinfo {pages} {245115}
  (\bibinfo {year} {2006})}\BibitemShut {NoStop}%
\bibitem [{\citenamefont {Li}\ \emph {et~al.}(2006)\citenamefont {Li},
  \citenamefont {Yu},\ and\ \citenamefont {Jin}}]{li2006}%
  \BibitemOpen
  \bibfield  {author} {\bibinfo {author} {\bibfnamefont {L.}~\bibnamefont
  {Li}}, \bibinfo {author} {\bibfnamefont {W.}~\bibnamefont {Yu}}, \ and\
  \bibinfo {author} {\bibfnamefont {C.}~\bibnamefont {Jin}},\ }\href@noop {}
  {\bibfield  {journal} {\bibinfo  {journal} {Phys. Rev. B}\ }\textbf {\bibinfo
  {volume} {73}},\ \bibinfo {pages} {174115} (\bibinfo {year}
  {2006})}\BibitemShut {NoStop}%
\bibitem [{\citenamefont {Blaha}\ \emph {et~al.}()\citenamefont {Blaha},
  \citenamefont {Schwarz}, \citenamefont {Madsen}, \citenamefont {Kvasnicka},\
  and\ \citenamefont {Luitz}}]{wien2k}%
  \BibitemOpen
  \bibfield  {author} {\bibinfo {author} {\bibfnamefont {P.}~\bibnamefont
  {Blaha}}, \bibinfo {author} {\bibfnamefont {K.}~\bibnamefont {Schwarz}},
  \bibinfo {author} {\bibfnamefont {G.~K.~H.}\ \bibnamefont {Madsen}}, \bibinfo
  {author} {\bibfnamefont {D.}~\bibnamefont {Kvasnicka}}, \ and\ \bibinfo
  {author} {\bibfnamefont {J.}~\bibnamefont {Luitz}},\ }\href@noop {}
  {}\bibinfo {note} {\texttt{Wien2K}, an augmented plane wave+local orbitals
  program for calculating crystal properties, Techn. Universit\"at Wien,
  Getreidemarkt 9/156A, 1060 Wien, Austria (2001).}\BibitemShut {Stop}%
\bibitem [{\citenamefont {Anisimov}\ \emph {et~al.}(1991)\citenamefont
  {Anisimov}, \citenamefont {Zaanen},\ and\ \citenamefont
  {Andersen}}]{anisimov1991}%
  \BibitemOpen
  \bibfield  {author} {\bibinfo {author} {\bibfnamefont {V.~I.}\ \bibnamefont
  {Anisimov}}, \bibinfo {author} {\bibfnamefont {J.}~\bibnamefont {Zaanen}}, \
  and\ \bibinfo {author} {\bibfnamefont {O.~K.}\ \bibnamefont {Andersen}},\
  }\href@noop {} {\bibfield  {journal} {\bibinfo  {journal} {Phys. Rev. B}\
  }\textbf {\bibinfo {volume} {44}},\ \bibinfo {pages} {943} (\bibinfo {year}
  {1991})}\BibitemShut {NoStop}%
\bibitem [{\citenamefont {Anisimov}\ \emph {et~al.}(1993)\citenamefont
  {Anisimov}, \citenamefont {Solovyev}, \citenamefont {Korotin}, \citenamefont
  {Czyzyk},\ and\ \citenamefont {Sawatzky}}]{anisimov1993}%
  \BibitemOpen
  \bibfield  {author} {\bibinfo {author} {\bibfnamefont {V.~I.}\ \bibnamefont
  {Anisimov}}, \bibinfo {author} {\bibfnamefont {I.~V.}\ \bibnamefont
  {Solovyev}}, \bibinfo {author} {\bibfnamefont {M.~A.}\ \bibnamefont
  {Korotin}}, \bibinfo {author} {\bibfnamefont {M.~T.}\ \bibnamefont {Czyzyk}},
  \ and\ \bibinfo {author} {\bibfnamefont {G.~A.}\ \bibnamefont {Sawatzky}},\
  }\href@noop {} {\bibfield  {journal} {\bibinfo  {journal} {Phys. Rev. B}\
  }\textbf {\bibinfo {volume} {48}},\ \bibinfo {pages} {16929} (\bibinfo {year}
  {1993})}\BibitemShut {NoStop}%
\bibitem [{\citenamefont {Chioncel}\ \emph {et~al.}(2007)\citenamefont
  {Chioncel}, \citenamefont {Allmaier}, \citenamefont {Arrigoni}, \citenamefont
  {Yamasaki}, \citenamefont {Daghofer}, \citenamefont {Katsnelson},\ and\
  \citenamefont {Lichtenstein}}]{chioncel2007}%
  \BibitemOpen
  \bibfield  {author} {\bibinfo {author} {\bibfnamefont {L.}~\bibnamefont
  {Chioncel}}, \bibinfo {author} {\bibfnamefont {H.}~\bibnamefont {Allmaier}},
  \bibinfo {author} {\bibfnamefont {E.}~\bibnamefont {Arrigoni}}, \bibinfo
  {author} {\bibfnamefont {A.}~\bibnamefont {Yamasaki}}, \bibinfo {author}
  {\bibfnamefont {M.}~\bibnamefont {Daghofer}}, \bibinfo {author}
  {\bibfnamefont {M.~I.}\ \bibnamefont {Katsnelson}}, \ and\ \bibinfo {author}
  {\bibfnamefont {A.~I.}\ \bibnamefont {Lichtenstein}},\ }\href@noop {}
  {\bibfield  {journal} {\bibinfo  {journal} {Phys. Rev. B}\ }\textbf {\bibinfo
  {volume} {75}},\ \bibinfo {pages} {140406(R)} (\bibinfo {year}
  {2007})}\BibitemShut {NoStop}%
\bibitem [{\citenamefont {Pavarini}\ \emph {et~al.}(2004)\citenamefont
  {Pavarini}, \citenamefont {Biermann}, \citenamefont {Poteryaev},
  \citenamefont {Lichtenstein}, \citenamefont {Georges},\ and\ \citenamefont
  {Andersen}}]{pavarini2004}%
  \BibitemOpen
  \bibfield  {author} {\bibinfo {author} {\bibfnamefont {E.}~\bibnamefont
  {Pavarini}}, \bibinfo {author} {\bibfnamefont {S.}~\bibnamefont {Biermann}},
  \bibinfo {author} {\bibfnamefont {A.}~\bibnamefont {Poteryaev}}, \bibinfo
  {author} {\bibfnamefont {A.~I.}\ \bibnamefont {Lichtenstein}}, \bibinfo
  {author} {\bibfnamefont {A.}~\bibnamefont {Georges}}, \ and\ \bibinfo
  {author} {\bibfnamefont {O.~K.}\ \bibnamefont {Andersen}},\ }\href@noop {}
  {\bibfield  {journal} {\bibinfo  {journal} {Phys. Rev. Lett.}\ }\textbf
  {\bibinfo {volume} {92}},\ \bibinfo {pages} {176403} (\bibinfo {year}
  {2004})}\BibitemShut {NoStop}%
\bibitem [{Note1()}]{Note1}%
  \BibitemOpen
  \bibinfo {note} {The hopping integral $V(e)$ between a Cr orbital of $e$
  symmetry and its O $2p$ ligands in a CrO$_4$ tetrahedral cluster is about
  $3.05-3.30$~eV according to LDA [$(pd\pi )=1.87-2.02$~eV], while the hopping
  integral $V(t_{2g})$ between a $t_{2g}$ orbital and its oxygen ligands in a
  CrO$_6$ octahedral cluster is substantially smaller, about 2.20~eV for
  LaCrO$_3$ [$(pd\pi )=1.10$~eV].}\BibitemShut {Stop}%
\bibitem [{\citenamefont {Maignan}\ \emph {et~al.}(2004)\citenamefont
  {Maignan}, \citenamefont {Caignaert}, \citenamefont {Raveau}, \citenamefont
  {Khomskii},\ and\ \citenamefont {Sawatzky}}]{maignan2004}%
  \BibitemOpen
  \bibfield  {author} {\bibinfo {author} {\bibfnamefont {A.}~\bibnamefont
  {Maignan}}, \bibinfo {author} {\bibfnamefont {V.}~\bibnamefont {Caignaert}},
  \bibinfo {author} {\bibfnamefont {B.}~\bibnamefont {Raveau}}, \bibinfo
  {author} {\bibfnamefont {D.}~\bibnamefont {Khomskii}}, \ and\ \bibinfo
  {author} {\bibfnamefont {G.}~\bibnamefont {Sawatzky}},\ }\href@noop {}
  {\bibfield  {journal} {\bibinfo  {journal} {Phys. Rev. Lett.}\ }\textbf
  {\bibinfo {volume} {93}},\ \bibinfo {pages} {026401} (\bibinfo {year}
  {2004})}\BibitemShut {NoStop}%
\bibitem [{\citenamefont {Chang}\ \emph {et~al.}(2009)\citenamefont {Chang},
  \citenamefont {Hu}, \citenamefont {Wu}, \citenamefont {Burnus}, \citenamefont
  {Hollmann}, \citenamefont {Benomar}, \citenamefont {Lorenz}, \citenamefont
  {Tanaka}, \citenamefont {Lin}, \citenamefont {Hsieh}, \citenamefont {Chen},\
  and\ \citenamefont {Tjeng}}]{chang2009}%
  \BibitemOpen
  \bibfield  {author} {\bibinfo {author} {\bibfnamefont {C.~F.}\ \bibnamefont
  {Chang}}, \bibinfo {author} {\bibfnamefont {Z.}~\bibnamefont {Hu}}, \bibinfo
  {author} {\bibfnamefont {H.}~\bibnamefont {Wu}}, \bibinfo {author}
  {\bibfnamefont {T.}~\bibnamefont {Burnus}}, \bibinfo {author} {\bibfnamefont
  {N.}~\bibnamefont {Hollmann}}, \bibinfo {author} {\bibfnamefont
  {M.}~\bibnamefont {Benomar}}, \bibinfo {author} {\bibfnamefont
  {T.}~\bibnamefont {Lorenz}}, \bibinfo {author} {\bibfnamefont
  {A.}~\bibnamefont {Tanaka}}, \bibinfo {author} {\bibfnamefont {H.-J.}\
  \bibnamefont {Lin}}, \bibinfo {author} {\bibfnamefont {H.~H.}\ \bibnamefont
  {Hsieh}}, \bibinfo {author} {\bibfnamefont {C.~T.}\ \bibnamefont {Chen}}, \
  and\ \bibinfo {author} {\bibfnamefont {L.~H.}\ \bibnamefont {Tjeng}},\
  }\href@noop {} {\bibfield  {journal} {\bibinfo  {journal} {Phys. Rev. Lett.}\
  }\textbf {\bibinfo {volume} {102}},\ \bibinfo {pages} {116401} (\bibinfo
  {year} {2009})}\BibitemShut {NoStop}%
\end{thebibliography}
%

\end{document}